\documentstyle[12pt]{article}
\font\tenbf=cmbx10
\font\tenrm=cmr10
\font\tenit=cmti10
\font\elevenbf=cmbx10 scaled\magstep 1
\font\elevenrm=cmr10 scaled\magstep 1
 1

\def\Re{{\cal R \mskip-4mu \lower.1ex \hbox{\it e}\,}}
\def\Im{{\cal I \mskip-5mu \lower.1ex \hbox{\it m}\,}}
\def\ie{{\it i.e.}}
\def\eg{{\it e.g.}}

\def\etal{{\it et al.}}
\def\ibid{{\it ibid}.}
\def\sub#1{_{\lower.25ex\hbox{$\scriptstyle#1$}}}
\def\sul#1{_{\kern-.1em#1}}
\def\sll#1{_{\kern-.2em#1}}
\def\sbl#1{_{\kern-.1em\lower.25ex\hbox{$\scriptstyle#1$}}}
\def\ssb#1{_{\lower.25ex\hbox{$\scriptscriptstyle#1$}}}
\def\sbb#1{_{\lower.4ex\hbox{$\scriptstyle#1$}}}

\def\tev{\,{\rm TeV}}
\def\gev{\,{\rm GeV}}
\def\mev{\,{\rm MeV}}
\def\to{\rightarrow}
\def\slash{\not\!}
\def\mh{\ifmmode m\sbl H \else $m\sbl H$\fi}
\def\mch{\ifmmode m_{H^\pm} \else $m_{H^\pm}$\fi}
\def\mt{\ifmmode m_t\else $m_t$\fi}
\def\mc{\ifmmode m_c\else $m_c$\fi}
\def\mz{\ifmmode M_Z\else $M_Z$\fi}
\def\mw{\ifmmode M_W\else $M_W$\fi}
\def\mws{\ifmmode M_W^2 \else $M_W^2$\fi}
\def\mhs{\ifmmode m_H^2 \else $m_H^2$\fi}
\def\mzs{\ifmmode M_Z^2 \else $M_Z^2$\fi}
\def\mts{\ifmmode m_t^2 \else $m_t^2$\fi}
\def\mcs{\ifmmode m_c^2 \else $m_c^2$\fi}
\def\mchs{\ifmmode m_{H^\pm}^2 \else $m_{H^\pm}^2$\fi}
\def\ztwo{\ifmmode Z_2\else $Z_2$\fi}
\def\zone{\ifmmode Z_1\else $Z_1$\fi}
\def\mtwo{\ifmmode M_2\else $M_2$\fi}
\def\mone{\ifmmode M_1\else $M_1$\fi}
\def\tb{\ifmmode \tan\beta \else $\tan\beta$\fi}
\def\xw{\ifmmode x\sub w\else $x\sub w$\fi}
\def\ch{\ifmmode H^\pm \else $H^\pm$\fi}
\def\lum{\ifmmode {\cal L}\else ${\cal L}$\fi}
\def\inpb{\ifmmode{\,{\rm pb}^{-1}}\else $\,{\rm pb}^{-1}$\fi}
\def\infb{\ifmmode{\,{\rm fb}^{-1}}\else $\,{\rm fb}^{-1}$\fi}
\def\epem{\ifmmode e^+e^-\else $e^+e^-$\fi}
\def\ppb{\ifmmode \bar pp\else $\bar pp$\fi}

\newskip\zatskip \zatskip=0pt plus0pt minus0pt
\def\matth{\mathsurround=0pt}
\def\lsim{\mathrel{\mathpalette\atversim<}}
\def\gsim{\mathrel{\mathpalette\atversim>}}
\def\atversim#1#2{\lower0.7ex\vbox{\baselineskip\zatskip\lineskip\zatskip
  \lineskiplimit 0pt\ialign{$\matth#1\hfil##\hfil$\crcr#2\crcr\sim\crcr}}}

\renewenvironment{thebibliography}[1]
 { \elevenrm
   \begin{list}{\arabic{enumi}.}
    {\usecounter{enumi} \setlength{\parsep}{0pt}
     \setlength{\itemsep}{3pt} \settowidth{\labelwidth}{#1.}
     \sloppy
    }}{\end{list}}

\begin{document}

\rightline{\vbox{\halign{&#\hfil\cr
&ANL-HEP-CP-93-68\cr
&August 1993\cr}}}

\begin{center}
\vglue 0.6cm
{\tenbf {\bf EXTENDED GAUGE SECTORS AT LINEAR COLLIDERS }
\footnote{Research supported by the
U.S. Department of
Energy, Division of High Energy Physics, Contract W-31-109-ENG-38.}
\footnote{Presented at the {\it Workshop on Physics and Experiments with
Linear \epem\ Colliders}, Waikoloa, Hawaii, April 1993}
\vglue 1.0cm
{\tenrm JoAnne L. Hewett \\}
\baselineskip=13pt
{\tenit High Energy Physics Division, Argonne National Laboratory, \\}
\baselineskip=12pt
{\tenit Argonne, Illinois 60439,\  USA\\}}
\vglue 0.8cm
{\tenrm ABSTRACT}
\vglue 0.3cm
{\rightskip=3pc
 \leftskip=3pc
 \tenrm\baselineskip=12pt
 \noindent
Signatures of extended gauge models at high energy \epem\ linear colliders
are summarized.}
\end{center}
\vglue 0.6cm
\baselineskip=14pt

The phenomenology of models with extended gauge symmetries is particularly
rich.  The existence of new gauge bosons is one of the hallmarks of
extended electroweak theories and their discovery would be a definitive signal
for physics beyond the Standard Model (SM).  However, extra gauge bosons
are not the sole signature of an extended gauge group.  These models
also contain new exotic fermions, which are required for anomaly cancellation,
as well as an enlarged Higgs sector, to facilitate the extended
symmetry breaking.  In addition, Supersymmetry may also be present,
particularly in Grand Unified Theories (GUTS), to solve the hierarchy problem
and to ensure the consistency of coupling constant unification with present
data\cite{susy}.  This talk will summarize the
potential of high energy \epem\ linear colliders to investigate the physics
of new gauge bosons, at mass scales both above and equal to the machine
center of mass energy, and to discover signals for exotic fermions.
The subjects of extended Higgs sectors and Supersymmetry will be left to other
speakers\cite{other}.

We begin with a brief review of several extended gauge models which
presently appear in the literature.  The most appealing set of enlarged
electroweak models are those which are based on GUTS,  examples being the
unifying groups $SO(10)$ and $E_6$.  In $E_6$ string-inspired models\cite{esix}
additional $Z$-bosons arise from the breaking chain
\begin{eqnarray}
E_6\to SO(10)\times U(1)_\psi &\to & SU(5)\times U(1)_\chi\times U(1)_\psi
\nonumber \\
&\to & SM \times U(1)_\theta \,,
\end{eqnarray}
where $U(1)_\theta$, which is a linear combination of $U(1)_\chi$ and
$U(1)_\psi$, remains unbroken at low energies ($\lsim 10\tev$).  The fermion
couplings of the extra $Z$-boson, $Z'$, in this model take the form
${g\over{2c_w}}\sqrt{5x_w/3}
(Q_\psi\cos\theta/\sqrt 6 - Q_\chi\sin\theta/\sqrt{10})$, where $\theta$
lies in the range $-90^\circ\leq\theta\leq 90^\circ$, $Q_{\psi,\chi}$
are determined by group theory, $x_w=\sin^2\theta_w$, and
$c_w=\cos\theta_w$.  Special cases in this category are
$\theta=0^\circ$ (model $\psi$), $\theta=-90^\circ$ (model $\chi$), and
$\theta=\arcsin(\sqrt{3/8})$ (model $\eta$).  The latter case represents the
rank-5 model derived directly from the flux breaking of superstring theories.

$SO(10)$ GUTS\cite{soten} leads to the intermediate symmetries, \eg,
\begin{eqnarray}
SO(10) &\to & SU(3)_C\times SU(2)_L\times U(1)_Y\times U(1)_\chi \nonumber \\
&\to & SU(3)_C\times SU(2)_L\times SU(2)_R\times U(1)_{B-L} \,.
\end{eqnarray}
The first chain leads to the additional $Z$-boson, $Z_\chi$, discussed above,
while the second chain yields right-handed charged as well as neutral
currents and is denoted as the left-right symmetric model (LRM).  In the LRM,
the $Z'$ couples to ${g\over {2c_w}}(\kappa-(1+\kappa)x_w)^{-1/2}[x_wT_{3L}+
\kappa(1-x_w)T_{3R}-x_wQ]$, with $0.55\leq \kappa\equiv g_R/g_L\leq 1-2$
\cite{mohap}, $T_{3L(R)}$ is the fermion left-(right-)handed isospin, and $Q$
is the fermion electric charge.  Note that for strict left-right symmetry,
$\kappa=1$.  Another extended model based on the second intermediate group
above is the alternative left-right symmetric model (ALRM) \cite{ma}, which
is embedded in $E_6$ GUTS and switches the quantum number assignments between
some of the ordinary fermions and exotic fermions contained in the
{\bf 27} of $E_6$.  In this case the right-handed $W$-boson carries lepton
number and has odd R-parity, thus avoiding\cite{ma} the usual constraints on
the mass of right-handed $W$'s.

There are, of course, many other models with extended electroweak sectors
that are not based on GUTS.  In order to get a feel for the variety of such
models, we list a few of them here.  For example, the sequential standard model
(SSM) contains a $Z'$ which is just a heavier version of the SM $Z$.  While
this model is lacking in theoretical motivation, it does provide a useful
benchmark in comparing experimental constraints and capabilities.  A recently
revived model\cite{suthree}, based on the gauge symmetry $SU(3)_C\times SU(3)
\times U(1)$, provides interesting production mechanisms at $e^-e^-$ colliders
which will be discussed further in this report.  The un-unified model
\cite{harv} extends the electroweak gauge group to $SU(2)_\ell\times SU(2)_q
\times U(1)_Y$, where the quark and lepton generations transform under their
own $SU(2)$.  Other models extend the color gauge group, such as that of Foot
and Hernandez \cite{oscar}, which is based on $SU(5)_C\times SU(2)_L\times
U(1)_{Y'}$.  There are several theories based on horizontal interactions,
such as the $S_{P}(6)_L\times U(1)$ model of Bagneid \etal \cite{kuo}.
In the generational model \cite{lima}, each generation transforms under its
own $SU(2)$.  And in the Leptophilic model \cite{xghe}, differences in
lepton number are gauged.

\vglue 0.6cm
{\elevenbf\noindent New Gauge Bosons \hfil}
\vglue 0.4cm
In all of the above models, the $Z-Z'$ mass matrix takes the form
\begin{equation}
{\cal M}^2 = \left( \begin{array}{cc}
                     M_Z^2  & \gamma M_Z^2  \\
                     \gamma M_Z^2  &  M_{Z'}^2
                     \end{array} \right)  \,,
\end{equation}
where $\gamma$ is determined within each model once the Higgs sector is
specified.  For example, $\gamma_{LRM}=-\sqrt{1-2x_w}$ (for $\kappa=1$),
and $\gamma_{E_6}=
-2\sqrt{5x_w/3}[(\cos\theta/\sqrt{6}-\sin\theta/\sqrt{10})\tan^2\beta
-(\cos\theta/\sqrt{6}+\sin\theta/\sqrt{10})](1+\tan^2\beta)^{-1}$, where
$\tan\beta$ is the ratio of vacuum expectation values (VEVs) of the two
Higgs doublets present in $E_6$ superstring models.  The physical eigenstates
are then
\begin{eqnarray}
Z_1 & = & Z'\sin\phi + Z\cos\phi \,, \nonumber  \\
Z_2 & = & Z'\cos\phi - Z\sin\phi \,,
\end{eqnarray}
where $Z_1$ is currently being probed at LEP, and $\tan 2\phi=2\gamma M_Z^2/
(M_Z^2-M_{Z'}^2)$ with $|\phi|\lsim 0.01$ from LEP (Ref.\ \cite{radcor}).

Precision measurements constrain\cite{radcor} extended gauge sectors, by
limiting the indirect \ztwo\ contributions to processes such as $\mu$-decay,
low-energy neutral currents, atomic parity violation, the $W$ mass, and the
properties of the $Z_1$ boson.  In Fig.\ 1, from Langacker and
Luo\cite{radcor},
the the bounds on $M_2$ are presented as a function of the $Z-Z'$ mixing angle
$\phi$ in $E_6$ models $\chi$ and $\eta$ for the case where $m_{top}$ is
unconstrained.  The allowed regions are enclosed by the curves and the
dotted curve represents the additional constraints imposed from the Higgs
sector as described in the above discussion of the $Z-Z'$ mass matrix.
It should  be noted that these limits are slightly sensitive to the effects
of other possible indirect contributions, such as, the top-quark, Higgs bosons,
supersymmetry, and exotic fermions.
Future improvements in these indirect bounds will be obtained as the data
becomes ever more precise and the top-quark is discovered.

Direct searches for new gauge bosons at hadron colliders are performed
via the production mechanisms, $p\bar p\to Z'\to \ell^+\ell^-$ and $p\bar p
\to W'\to \ell^\pm\slash p_T$.  From the 1988-89 Tevatron
run with 4.7 \inpb, CDF has set
a $95\%$ C.L. bound\cite{cdfztwo} on the \ztwo\ mass of 412 (320, 340, 340,
310) GeV in the SSM (Models $\psi$, $\chi$, $\eta$, and the LRM, respectively),
using both $e$ and $\mu$ data.  A limit has also been set on a possible new
$W'$ with SM strength couplings of 520 GeV.  The recent 1992-93 run with
approximately $21-22$ \inpb\ of integrated luminosity yield\cite{cdfssi} the
{\it very} preliminary $95\%$ C.L. constraint of $M_2 > 495\gev$ in
the SSM from the CDF electron data alone.
Existing and future bounds obtainable at the Tevatron on new
gauge bosons in $E_6$ models are shown in Fig.\ 2a as a function of $\theta$
for various values of integrated luminosity.  These calculations assume
$e + \mu$ data samples with CDF detector efficiencies, and include a 2-loop
K-factor as well as a finite top-quark mass.  Assuming 400 \inpb, we see that
the Tevatron will be probing \ztwo\ masses of order 700-800 GeV in these
models.
At the Tevatron, one may also search for new gauge bosons in the
$Z', W'\to 2$ jets channel, providing the large QCD background can be overcome.
The dijet invariant mass spectrum at the Tevatron can give\cite{tgrdijet}
additional bounds on the mass of new $W$-bosons, depending on their coupling
strength, but the lepton data is found to give the best constraints on most
new neutral gauge bosons.  Hadron supercolliders, \ie, the SSC and LHC,
can search for \ztwo\ bosons with masses up to 3-7 TeV, depending on the
model.  For example, Fig.\ 2b displays the discovery limit as a function
of integrated luminosity for a \ztwo\ from the LRM at the SSC (solid
curve) and LHC (dashed curve), assuming a $5\sigma$ signal in the $e$
channel only and $\kappa=1$.  At design luminosity, the SSC(LHC) could
detect a LRM \ztwo\ with $M_2\leq 6.1 (4.9) \tev$.
We note that in the search regions presented here, it is assumed that the
\ztwo\ decays only into 3 generations of SM fermions.

If a \ztwo\ is discovered at a hadron collider, a much more difficult
puzzle develops -- the identification of the extended electroweak model
from which the \ztwo\ originates.  Numerous studies of this issue have been
performed during the last few years and have been summarized at this
meeting by Cveti\` c\cite{mirjam}.  Several ideas has been proposed, including
$Z_2+\gamma,Z,W$ associated production, leptonic forward-backward and final
state tau polarization asymmetries,
3-body \ztwo\ decays, \ztwo\ production with polarized beams, and
extracting a signature from the
$Z_2\to 2$ jet channel.  All of the techniques proposed thus far (with the
exception of the leptonic forward-backward asymmetry) suffer from at least
one of the following problems,
(i) the event rate dies off rapidly unless the
\ztwo\ mass is in the range 1-2 TeV, and (ii) it is difficult to extract the
signal from the overwhelming background.

LEP II and future \epem\ colliders allow for indirect searches in the case
$\mtwo>\sqrt s$ by looking for possible deviations from SM expectations for
cross sections and asymmetries associated with the reaction $\epem\to f\bar f$.
This is similar in concept to exploring modifications of QED predictions due to
the SM $Z$ at PEP/PETRA energies.  The limits obtained by this method are
very model-dependent and are quite sensitive to the integrated luminosity,
the value of $\sqrt s$, as well as the flavor of the final state fermion.
The measurable quantities associated with $\epem\to f\bar f$ that are sensitive
to \ztwo\ exchange are (i) total cross section, $\sigma^f$, (ii) the ratio
of hadronic to leptonic cross section, $\sigma^{had}/\sigma^\ell$, (iii) the
forward-backward asymmetry, $A_{FB}^f$, (iv) final state polarization
asymmetry of taus, $A_{pol}^\tau$, and if longitudinal polarized beams are
available, (v) the left-right asymmetry, $A_{LR}^f$, and (vi) the polarized
forward-backward asymmetry, $A_{FB}^f(pol)$.  As an example of the power of
this search technique,
Fig.\ 3a presents the possible bounds on \mtwo\ in $E_6$ models as
a function of $\theta$ from a $1\sigma$ deviation in $\delta\sigma/\sigma$
for $\mu, c,$ and $b$ final states with 200 \inpb\ of integrated
luminosity at LEP II.  Note that the discovery region extends to values of the
\ztwo\ mass that are $2-3\times\sqrt s$ and is comparable to the future search
reach of the Tevatron.  A $\chi^2$ analysis has been performed in Ref.\
\cite{finland} for a $\sqrt s=500\gev$ linear collider, where the above
processes (i) and (iii)--(vi) have been examined, including 3-loop QCD
corrections, oblique electroweak corrections, identification efficiencies of
final state particles, and a beam polarization measurement error of
$\delta P/P=1\%$.  Figure 3b displays the $95\%$ C.L. search limits from
this study for a \ztwo\ arising from $E_6$ models as a function of the model
parameter $\theta$ assuming ${\cal L}=10$ or $25\infb$ with or without beam
polarization ($P=90\%, 0$).  In general, the discovery region reaches
$3-6\times\sqrt s$ for ${\cal L}=25\infb$.  The results of a separate
investigation\cite{djouadi} are shown in Fig.\ 4 for a \ztwo\ originating from
(a) $E_6$ models as a function of $\cos\beta$ where $\beta=\theta+90^\circ$ and
(b) LRM as a function of $\alpha_{LR}$ (where $\alpha_{LR}$ is related to
$\kappa$) for various values of the
center of mass energy.  Here, QCD, electroweak corrections, initial state
radiation, as well as final state identification efficiencies
are included for the processes (i)--(iii) and (v)
above.  While high-energy \epem\ colliders do have a large discovery reach
for extra gauge bosons, it is clear from these results that a $\sqrt s\sim
1\tev$ machine is needed in order to be competitive with the hadron
supercolliders.

How well does an \epem\ collider perform in determining the model of
origin of a new neutral gauge boson?  In Fig.\ 5a-c from \cite{finland}
a $\chi^2$ distribution as a function of the $E_6$ model parameter $\theta$
is presented assuming a \ztwo\ from Model $\psi$ ($\theta=0^\circ$) with
$M_2=1-2\tev$ is present in the data at $\sqrt s=500\gev$.  This analysis
examines the processes (i) and (iii)--(vi) above with an integrated
luminosity of 5 or 25 \infb\ and
both polarized ($P=90\%$) and unpolarized beams.  The `measured' range of
$\theta$ at $95\%$ C.L. corresponds to the points where the solid horizontal
line intersects the parabolic curves.  A complimentary analysis\cite{djouadi}
is shown in Fig.\ 5 (d-f), where the ability to discriminate between a
\ztwo\  from $E_6$ Models
and one from the LRM is examined.  Here, the unshaded areas represent
the regions in the $\alpha_{LR} - \beta$ plane where
the two models are distinguishable and the single-hatched area corresponds
to the region of confusion,  using the quantities (i)--(iii) above.  If one
includes polarized beams and process (v), then the area of confusion is
reduced to the double-hatched region.  It is clear from these studies that
\epem\ machines are capable of determining the couplings of a relatively
light \ztwo\ ($\sim 1\tev$)!

The existence of new gauge bosons may also modify the reaction $\epem\to
W^+W^-$, which is notably sensitive to the specific form of the gauge
couplings.   The \ztwo\ participates in this process only through $Z-Z'$
mixing and hence $\epem\to W^+W^-$ can be used as a probe of the amount
of mixing present, \ie, the value of $\phi$.
Here, we consider two observables, the total cross section and
the polarized left-right asymmetry.  The $95\%$ C.L. search reach
at a 1 TeV collider with 100 \infb\ of integrated luminosity in
$E_6$ models with $M_2>\sqrt s$ is shown in Fig.\ 6a as a function of the
parameter $\theta$  for various values of $\tan\beta$.  (Recall
that the $Z-Z'$ mixing angle $\phi$ is determined in $E_6$ models once the
\zone\ and \ztwo\ boson masses, $\theta$ and $\tan\beta$ are known.)
The relatively small search limits reflect the fact that $\phi$ is tightly
constrained by LEP data, and the regions where is the search limit disappears,
for example at $\theta=0^\circ$ with $\tb=1$, is due to the fact that
$\phi=0$ close to these values of the parameters.  The \ztwo\ discovery window
from measurements of the left-right asymmetry is, of course, quite sensitive
to the amount of beam polarization and its associated error.  The anticipated
error in $A_{LR}$ due to finite polarization is
\begin{equation}
\delta A_{LR}=\left[ {1-(PA_{LR})^2\over N_WP^2}\oplus
\left( {A_{LR}\delta P\over P}\right)^2 \right]^{1/2} \,,
\end{equation}
where $N_W=\sigma{\cal L}\epsilon_W$ with $\epsilon_W$ being the $W$
identification efficiency, and $`\oplus'$ represents that the errors are
added in quadrature.  The $95\%$ C.L. search reach  from $A_{LR}$
and $\sigma$ combined is presented in Fig.\ 6b for the LRM as a function
of $\kappa$, with $P=90\%~{\rm or}~100\%$ and $\delta P/P=0~{\rm or}~1\%$.
We see that as soon as the uncertainty in the amount of polarization is
included, most of the sensitivity to \ztwo\ exchange is lost.
$\epem\to W^+W^-$ may be more interesting in the case where the \ztwo\ is
on resonance, which will be discussed below.

In principle, new neutral gauge bosons will contribute to Bhabha
and Moller scattering, see \eg, Ref.\ \cite{esix}.  However, in practice
these processes are dominated by the $\gamma$-pole and are not very
sensitive to indirect \ztwo\ contributions below production threshold.

Now, we turn our attention to on-resonance \ztwo\ production at a
TeV \epem\ collider.  We note that most of the discussion in the
literature\cite{mirjam} on the production and model identification of
\ztwo\ bosons at hadron supercolliders has focused on the $1-2\tev$ mass
range, and clearly, the discovery of a $\sim 1\tev$ \ztwo\ at the SSC/LHC
(which will presumably, but not necessarily, be built before
a high-energy \epem\ linear accelerator!), would provide powerful motivation
for building a TeV \epem\ machine.  The event rates are, of course, quite large
on-resonance, yielding enhancements of $2-3$ orders of magnitude over the
usual falling SM cross sections.  Such a machine would become a \ztwo\ factory
in a manner similar to the \zone\ production at LEP I.  In this case,
unraveling
the various couplings of the \ztwo\ and determining its model of origin
would become a much easier puzzle to solve.  The \ztwo\ properties which can
be measured accurately (besides its mass) include (i) the \ztwo\ partial widths
into all identifiable final states, $\Gamma_\ell, \Gamma_b, \Gamma_t,$ and
$\Gamma_{had}$, as well as perhaps $\Gamma(\ztwo\to$ exotics, superpartners)
if these channels are available.
(ii) The total width, $\Gamma_{tot}$, which could also yield information on
possible invisible decay channels.  (iii)  Universality should be verified,
as it is violated in some extended models.  (iv) Several asymmetries can
be determined for every identifiable final state.  For $f=\ell,b,~{\rm or}~t$,
we have
\begin{eqnarray}
A_{FB}^f & = & 3A_eA_f/4 \,, \nonumber \\
A_{FB}^f(pol) & = & 3A_f/4 \,, \\
A_{LR}^\ell & = & A_e = A_{pol}^\tau \,, \nonumber
\end{eqnarray}
with the definition $A_i=2v_ia_i/(v_i^2+a_i^2)$.  The forward-backward
asymmetry for the process $\epem\to\mu^+\mu^-$ is presented in Fig.\ 7a as
a function of the $E_6$ parameter $\theta$, including the statistical errors
calculated for ${\cal L}=25\infb$ at $\sqrt s=1\tev$.  The values of
$A_{FB}^\mu$ in the LRM (with $\kappa=1$), the SSM, and the ALRM are also
shown for comparison.  We see from the figure that although this asymmetry
can be well measured, it alone can not uniquely differentiate between the
various models.

Some extended electroweak models possess specific \ztwo\ properties  that are
unique to a new neutral gauge boson originating within that particular
scenario.
For example, in $E_6$ theories, the \ztwo-fermion vector and axial vector
coupling constants obey the relations $v_\ell=-v_b$ and $a_\ell=a_b$, which
yields the predictions
\begin{eqnarray}
&{{\Gamma(\ztwo\to b\bar b)}\over {\Gamma(\ztwo\to\ell^+\ell^-)}}
= 3\oplus {\rm QCD~corrections}\,, \nonumber \\
& A_{FB}^\ell=-A_{FB}^b \,, \\
& A_{FB}^b(pol)=-{3\over 4}A_{LR}^\mu \,, \nonumber
\end{eqnarray}
while the $E_6$ property $v_{u,c,t}=0$ implies
\begin{eqnarray}
A_{FB}^t  = 0 & = & A_{FB}^t(pol) \,, \\
              & = & A_{LR}^t \,. \nonumber
\end{eqnarray}
These relations are shown explicitly in Fig.\ 7b where the forward-backward
asymmetry is presented for $\mu, b,~{\rm and}~t$ final states as a function
of $\theta$.  In the case of the LRM, the forward-backward asymmetry for
$\mu, b$ and $t$ final states as well as $A_{LR}^\mu$ vanish at two
distinct values of $\kappa$, $\kappa=\kappa_{min}=\sqrt{x_w/(1-x_w)}$ and
$\kappa=\sqrt{3x_w/(1-x_w)}$.  The forward-backward asymmetry and left-right
asymmetry for $\mu, b$ and $t$ final states and the ratio $\Gamma_{b,t}/
\Gamma_\ell$ are shown in Fig.\ 8a--c for the LRM as a function of $\kappa$.
These figures clearly display that the asymmetries do indeed vanish for these
two values of $\kappa$.  The results presented here are essentially unmodified
by small values of $Z-Z'$ mixing and finite values of $m_t/\mtwo$.
These distinctive properties make excellent on-resonance tests of $E_6$
and left-right symmetric models!

As a further example of the power of these on-resonance tests, we show
the areas of the $\Gamma_t/\Gamma_\mu - \Gamma_b/\Gamma_\mu$ and
$\Gamma_b/\Gamma_\mu - A_{FB}^\mu$ planes that are populated by three
different extended models in Fig.\ 9a--b, respectively.  In each case, the
solid curve represents $E_6$ models as the parameter $\theta$ is varied,
the dotted curve corresponds to the LRM as $\kappa$ is varied, and the
dashed curve is the un-unified model as its model parameter ($\sin\phi$,
which represents the relative amount of mixing between the $SU(2)_\ell$ and
$SU(2)_q$ factors).  Note that these three models cover quite distinct
regions in these planes.

As discussed above, the decay $\ztwo\to W^+W^-$ can be a useful probe of
$Z-Z'$ mixing as well as the Higgs sector.
The total cross section for $\epem\to W^+W^-$ in the $E_6$
model $\chi$ with $M_2=1\tev$ (dashed curve) contrasted to that of the SM
(solid
curve) as a function of $\sqrt s$ is shown  in Fig.\ 10, from Ref.\
\cite{paver}, for $\Delta M\equiv M_Z-M_1 = 50\mev$.  Clearly, the existence
of a \ztwo\ greatly enhances the event rate;
for more details see \cite{esix,paver}.

\vglue 0.6cm
{\elevenbf\noindent New Fermions \hfil}
\vglue 0.4cm

There is a wealth of phenomenology associated with exotic fermions that are
present in extended gauge models, particularly in the case of $E_6$ theories,
and in the case where mirror fermions are present.
So much research has been performed in this area\cite{esix,revs,exferms}
that it could well be the subject of a separate review.  Here, we will restrict
ourselves to work which was presented at this meeting and related material.

In $E_6$ models, each chiral generation of fermions is assigned to the
{\bf 27} dimensional representation which contains the usual {\bf 16} of
$SO(10)$ (the SM fermions plus the right-handed $\nu$) as well as 11 new
fields.
Since the superstring-inspired $E_6$ theories are supersymmetric, the
corresponding scalar superpartners to these fermions are also present.
For each family one then has the additional fermions,
\begin{equation}
\left( \begin{array}{c}
N \\
E
\end{array} \right)_L  \,, \, \,
\left( \begin{array}{c}
N \\
E
\end{array} \right)^c_L  \,, \, \,
h_L \,, \, \, h^c_L \,, \, \, S^c_L \,,
\end{equation}
where $(N, E)$ are color singlet, vector-like leptons with $Q_N=0$ and
$Q_E=-1$,
h is a color triplet, iso-singlet fermion with $Q_h=-1/3$, and $S^c$ is a
neutral color and iso-singlet.
The most general superpotential allows for three possible
baryon and lepton number assignments for $h$, however, all terms in
the superpotential can not be simultaneously present or low-energy baryon
and lepton number violation will occur.   Thus, $h$ is forced to be either
a quark, a diquark, or a leptoquark.

These exotic fermions may be pair produced at \epem\ colliders via s-channel
$\gamma, \zone,$ and \ztwo\ exchange.  Clean signatures are expected up to
the kinematical limit, $m_F \simeq \sqrt s/2$.  A thorough background
study\cite{vpion} of $\epem\to N\bar N$ has verified that this holds true
for heavy neutrino production.   If the pair production of new
fermions is observed, then measurement of their forward-backward and left-right
asymmetries can be used as a probe\cite{oldjlh} of their electroweak
properties.
Possible mixing between the exotic and ordinary SM fermions can induce
single exotic production\cite{esix,abel} via $\epem\to E\bar e + \bar Ee,
N\bar\nu_e + \bar N\nu_e, h\bar d + \bar hd$.  These processes are mediated
through s-channel \zone\ and \ztwo\ exchange, as well as t-channel $Z_{1,2}(W)$
exchange in the case of $E\bar e(N\bar\nu_e)$ production, and are directly
proportional to the degree of ordinary-exotic mixing.  Present data
restricts\cite{radcor} this mixing to values $\theta_{mix}\lsim 0.1$.  This
yields cross sections\cite{abel} in the $10-1000\infb$ range for $m_F
<450-475\gev$ for $E\bar e$ or $N\bar\nu_e$ production with maximal mixing at
a 500 GeV machine (with the neutral cross section being larger than that for
the charged leptons).  The cross sections without the t-channel contribution
(\ie, single $h$ production) are somewhat smaller.  Azuelos and
Djouadi\cite{abel} have performed a Monte-Carlo study and have considered
several
background sources for both neutral and charged single lepton production.
These authors found that the ratio of signal events to the square root of the
background is greater then unity ($S/\sqrt B >1$) for mixing angles
$\theta_{mix}\gsim 0.005 (0.03)$ with $m_F=350\gev$ for neutral(charged)
lepton single production at $\sqrt s=500\gev$ and ${\cal L}=50\infb$.

Leptoquarks appear naturally in theories which place quarks and leptons
on equal footing, including $E_6$ models as mentioned above.  They
couple to a lepton-quark pair, with an {\it a priori} unknown
strength governed by the Yukawa coupling, $\lambda$.
For calculational purposes, these Yukawa couplings are usually parameterized
by $\lambda^2/4\pi=F\alpha_{em}$.  These particles can be light $\sim 100\gev$
and still avoid conflicts\cite{lq} with rapid proton decay and dangerously
large
flavor changing neutral currents.  This is particularly true in models
where each generation of fermions has its own leptoquark(s) which couples
only within that generation.  Present experimental bounds\cite{exptlq} on
scalar leptoquarks are (i) $M_{LQ}>116\gev$ at $95\%$ C.L. if
$B(LQ\to eu)=100\%$ from $gg, q\bar q\to LQ\ \bar{LQ}$ production at CDF,
and (ii) $M_{LQ}\gsim 175\gev$ at $95\%$ C.L., where $\lambda$ takes on values
equivalent to the electroweak coupling
strength, from the production $eq\to LQ$ at HERA.

Leptoquarks may also be pair produced and observed up to the kinematic limit
in \epem\ collisions.  For this case, QCD as well as QED (initial state
radiation and beamstrahlung) corrections have been recently
computed and reported at this meeting\cite{blum}.
The angular distributions in leptoquark pair production
are affected by the presence of the t-channel quark exchange diagram, and
are hence sensitive to the Yukawa coupling.  This  can be seen explicitly in
Fig.\ 11 from Bl\" umlein and R\" uckl\cite{blum}, where it is clear that
the angular distributions can be used to distinguish between scalar and vector
leptoquark production as well as between the existence of right- and
left-handed couplings.  Single production\cite{singlelq,leptos} occurs via
the reaction $e\gamma\to LQ+$ jet and is proportional to the value of the
Yukawa coupling.  Cross sections at $\sqrt s=500\gev$
for scalar leptoquarks using three
different photon sources, (i) Weizs\" acker-Williams distribution, (ii)
beamstrahlung, and (iii) backscattered laser beam, are displayed in Fig.\ 12a
from Ref.\ \cite{singlelq} (including a $p_T$ cut on the associated jet of
10 GeV).  Figure 12b shows the  discovery region\cite{singlelq}  from this
mechanism, demanding a $5\sigma$ signal for both
10 and 25 \infb\ of integrated luminosity.  We see that with large values of
the coupling strength discovery up to the kinematic limit is possible,
rendering this process competitive with the pair production process.

If leptoquarks are too heavy to be produced directly, perhaps they can be
detected through indirect effects.  In principle, leptoquarks can participate
in the reaction $\epem\to q\bar q$  via u- and t-channel exchanges, due to
the presence of the Yukawa couplings, and can produce deviations from the
SM predictions for cross sections and asymmetries. The $95\%$ C.L. bounds
which can be obtained at a 500 GeV \epem\ collider from such reactions is
presented in Fig.\ 13 in the leptoquark coupling strength - mass plane for
various values of integrated luminosity.  Here, the search region corresponds
to the area above the curves, and has been computed for scalar leptoquarks.

\vglue 0.6cm
{\elevenbf\noindent $e^-e^-$ Collisions \hfil}
\vglue 0.4cm

An interesting reaction in $e^-e^-$ collisions, a possible option at the next
linear \epem\ collider, is inverse neutrinoless
double beta decay, first proposed by Rizzo\cite{invbeta}.
Extended electroweak models with  Majorana neutrinos and heavy iso-singlet
neutral leptons can mediate $\Delta L=2$ interactions; at low-energies
such models can be probed indirectly by searching for rare processes
such as neutrinoless double $\beta$-decay.  High energy $e^-e^-$ collisions
may provide a new window into the $\Delta L=2$ sector of these models via
the reaction $e^-e^-\to W_i^-W_i^-$, where the $W_i$ may represent either the
SM $W_L$ boson or an additional charged gauge boson, such as the right-handed
$W_R$ of the LRM.  In the case where two SM
$W^-_L$ bosons are produced, there is a danger of unitarity violation, which
can
only be cured if either extra neutral fields plus mixing are included, or
a doubly charged Higgs boson ($\Delta$) is exchanged in the  s-channel.
The cross section in the former case is presented in Fig.\ 14a as a function
of the heavy neutrino mass with $\sqrt s = 500$ or 1000 GeV from
Rizzo\cite{invbeta}.  Note that these
results should be scaled by four powers of the light-heavy neutrino mixing
angle, which is expected to be $\lsim {\cal O}(10^{-2})$.

This $\Delta L=2$ reaction is quite natural in the LRM, as this model contains
all the necessary ingredients for the preservation of unitarity:
heavy right-handed neutrinos, a doubly-charged Higgs scalar,  as well as
right-handed $W_R$ bosons.  In the LRM case, the sum of the asymptotic
$\nu$ and $\nu^c$ contributions no longer cancel and a $\Delta$ exchange is
required in order to restore unitarity.  Here, $d\sigma|_{s\to\infty}
\sim m_N$, where $N$ represents the heavy right-handed neutrino, and
unitarity is maintained as long as $N$ (and $\Delta$) have a mass less
than $\sim$ 2 (10) TeV.  Generally the cross section for $e^-e^-\to
W^-_RW^-_R$ can be quite large, as is depicted in Fig.\ 14b as a function of
$\sqrt s$ from Maalampi \etal\cite{invbeta}.  In this figure the mass of the
right-handed
$W_R$ is taken to be 500 GeV, $M_\Delta=500\gev$ and $m_N=0.5,1,1.5\tev$.
We see that the event rate can be sizable for integrated luminosities in the
100 \infb\ range.
Mixing in the $\nu-\nu^c$ mass matrix does not significantly contribute to
$W^-_RW^-_R$ production, but can induce the mixed final state $W^-_LW^-_R$.
In the absence of $W_L-W_R$ mixing, this process will proceed via t- and
u-channel exchanges, but still obeys unitarity due to the opposite helicity
structures at the two vertices.  The cross section (from Rizzo\cite{invbeta})
for $e^-e^-\to W^-_LW^-_R$,
is displayed in Fig.\ 14c as a function of the mass
of the heavy neutrino, taking $m_{W_R}=480\gev$ and $\kappa=0.9$ and $\sqrt s=
1\tev$.  Here the result must be multiplied by two powers of the scaled
mixing angle, $\theta_{mix}/0.01$.  We see that this cross section is
reasonably small, but might be observable depending on the value of the mixing.

A second interesting possibility in $e^-e^-$ collisions is dilepton
production. As has been emphasized by Frampton\cite{suthree,paul}, dileptons
naturally appear in various extended electroweak models, including $SU(15)$
GUTS and the $SU(3)\times U(1)$ model discussed in the introduction.
Here, dileptons contribute in s-channel exchange, yielding a distinctive
resonance peak as shown in Fig.\ 15 from Ref.\ \cite{paul}
for a dilepton mass of 500 GeV.
\vglue 4.0in
{\elevenbf\noindent Conclusions \hfil}
\vglue 0.4cm

In summary, we see that extended gauge sectors yield an abundance of exciting
phenomenology at high energy \epem\ colliders!  The present bounds on
additional
gauge bosons are becoming stronger everyday, with mass limits being in the
several hundred GeV range.  Hadron supercolliders can discover \ztwo\ bosons
with masses up to several TeV, but will have more difficultly determining
the extended model from which the \ztwo\ originates.  \epem\ colliders can
indirectly probe the existence of new neutral gauge bosons up to
$M_2 \sim (3-6)\times\sqrt s$ for integrated luminosities ${\cal L}\simeq
25\infb~[\sqrt s/500\gev]^2$.  The possible discovery of a `light' \ztwo\
($\sim 1\tev$) at a hadron supercollider would provide an impetus for the
construction of a TeV \epem\ linear collider, as \ztwo\ on-resonance
physics has an overwhelming potential.  In the case of exotic fermions, there
are many exciting production mechanisms and signatures which are unique to
\epem\ collisions.  And, last, but not least, one should keep the $e^-e^-$
collider option open as a possible technique for finding striking and
unique signatures for new physics.

\vspace{0.6cm}
{\elevenbf \noindent Acknowledgements \hfil}
\vglue 0.4cm
I would like to thank the organizers for creating a stimulating (and
Hawaiian!) atmosphere at this meeting.
This work has been supported in part by the U.S. Department of Energy,
Division of High Energy Physics under contract W-31-109-ENG-38, and in part
by the Texas National Research Laboratory Commission.
\vglue 0.5cm
{\elevenbf\noindent  References \hfil}
\vglue 0.4cm
\def\MPL #1 #2 #3 {Mod.~Phys.~Lett.~{\bf#1},\ #2 (#3)}
\def\NPB #1 #2 #3 {Nucl.~Phys.~{\bf#1},\ #2 (#3)}
\def\PLB #1 #2 #3 {Phys.~Lett.~{\bf#1},\ #2 (#3)}
\def\PR #1 #2 #3 {Phys.~Rep.~{\bf#1},\ #2 (#3)}
\def\PRD #1 #2 #3 {Phys.~Rev.~{\bf#1},\ #2 (#3)}
\def\PRL #1 #2 #3 {Phys.~Rev.~Lett.~{\bf#1},\ #2 (#3)}
\def\RMP #1 #2 #3 {Rev.~Mod.~Phys.~{\bf#1},\ #2 (#3)}
\def\ZPC #1 #2 #3 {Z.~Phys.~{\bf#1},\ #2 (#3)}
\def\IJMP #1 #2 #3 {Int.~J.~Mod.~Phys.~{\bf#1},\ #2 (#3)}

\newpage

{\tenrm

\noindent
{Fig. 1.  The $90\%$ C.L. allowed region in the $M_2-\phi$ plane in (a)
Model $\chi$ and (b) Model $\eta$ from precision data as described in
the text.  $\rho_0$ is the tree-level $\rho$ parameter, and the
dotted curves represent the constraints from the minimal Higgs sector for
various values of the ratios of VEVs.}

\noindent
{Fig. 2.  (a) Existing and expected future search limits at the Tevatron for
new $Z$ bosons in $E_6$ models as a function of $\theta$ for various values
of integrated luminosity as shown.  (b)  Discovery limit as a function of
integrated luminosity at the SSC (solid curve) and LHC (dashed curve)
in the LRM.}

\noindent
{Fig. 3. Indirect \ztwo\ search limits as a function of the $E_6$ parameter
$\theta$ at (a) LEP II with 200 \inpb\ of integrated luminosity
using $\delta\sigma/\sigma$ alone with $\mu, b$ and $c$ final states,
(b) at $\sqrt s=500\gev$, assuming ${\cal L}=10\infb$ with $P=0$ (dots)
and $P=90\%$ (dashes) and ${\cal L}=25\infb$ with $P=0$ (dashed-dots)
and $P=90\%$ (solid).}

\noindent
{Fig. 4.  \ztwo\ discovery limits in (a) $E_6$ models and (b) LRM as a
function of the model parameters $\cos\beta$ and $\alpha_{LR}$, respectively
for $\sqrt s=$ 190 GeV with ${\cal L}=500\inpb$ (dashed-dotted),
500 GeV (solid), 1 TeV (dashed), and 2 TeV (dotted) with ${\cal L}=20\infb$.
In each case, the upper thin (lower thick) curves are with (without) beam
polarization.}

\noindent
{Fig. 5.  $\theta$ determination for $E_6$ model $\psi$ assuming (a)
$M_2=1\tev$, (b) $M_2=1.5\tev$, and (c) $M_2=2\tev$.  The inner (outer) pair
of dashed ($P=90\%$) and dotted ($P=0$) curves correspond to ${\cal L}=25
(5)\infb$.  The $95\%$ C.L. is determined when the horizontal solid line
intersects the parabolic curves.  $95\%$ C.L.
distinction between $E_6$ and LRM \ztwo\
bosons for (d) $M_2=1.5\tev$, (e) $M_2=2\tev$, and (f) $M_2=2.5\tev$ with
$\sqrt s=500\gev$ with ${\cal L}=20\infb$.  The single (double) hatched regions
denotes the area of confusion without (with) the availability of
polarized beams.}

\noindent
{Fig. 6. $95\%$ C.L. \ztwo\ search limits from the reaction $\epem\to W^+W^-$
with $sqrt s=1\tev$ and ${\cal L}=100\infb$. (a) In $E_6$ models as a function
of $\theta$ with $\tan\beta=1 (2, 10)$ corresponding to dotted (dashed,
solid) curve, using $\sigma$ alone, (b) In the LRM as a function of $\kappa$,
using $\sigma$ and $A_{LR}$, with $P=100\%\, \delta P/P=0$ (solid),
$P=90\%\, \delta P/P=0$ (dashed), and $P=90\%$ or $100\%\, \delta P/P=0.01$
(dashed-dotted).}

\noindent
{Fig. 7.  On-resonance forward-backward asymmetry
in $E_6$ models as a function of $\theta$, assuming $\sqrt s=M_2=1\tev$.
(a) In $\epem\to\mu^+\mu^-$, where
the statistical errors are displayed for ${\cal L}=25\infb$.
$A_{FB}^\mu$ is also shown in the LRM, SSM, and ALRM, for comparison.
(b) For $\mu$ (solid), $t$ (dotted), and $b$ (dashed) final states.}

\noindent
{Fig. 8.  On-resonance tests of the LRM as a function of $\kappa$ for final
states $\mu$ (solid), $t$ (dotted), $b$ (dashed) with $\sqrt s=1\tev$.
(a) The forward-backward asymmetry, (b) left-right asymmetry, and (c) the
ratio of partial widths $\Gamma_q/\Gamma_\ell$.}

\noindent
{Fig. 9.  On-resonance model identification, where the regions populated by
$E_6$ models as $\theta$ is varied (solid), LRM as $\kappa$ is varied
(dotted), and the Un-unified model as $\sin\phi$ is varied (dashed) are
displayed in the (a) $\Gamma_t/\Gamma_\mu - \Gamma_b/\Gamma_\mu$  and (b)
$\Gamma_b/\Gamma_\mu - A_{FB}^\mu$ planes.}

\noindent
{Fig. 10.  The total $\sigma(\epem\to W^+W^-)$ as a function of $\sqrt s$
for the $E_6$ model $\chi$ with $M_2=1\tev$.  The solid line represents the
SM behavior.}

\noindent
{Fig. 11.  Angular distributions of a scalar ($S_1$) and vector ($U_1$)
leptoquark with mass $M_{LQ}=400\gev$ and $\sqrt s=1\tev$ and the couplings
$\lambda_L=\lambda_R=0$ (solid), $\lambda_L/e=0.3$ and $\lambda_R=0$
(dashed), $\lambda_L=0$ and $\lambda_R/e=1$ (dotted).}

\newpage

\noindent
{Fig. 12.  (a) Cross section for single leptoquark production via
$e\gamma\to LQ+$ jet as a function of the leptoquark mass at $\sqrt s=500\gev$.
The various photon sources are Weizs\" acker-Williams (solid), beamstrahlung
(dashed), and backscattered laser beam (dashed-dotted), with the upper
(lower) curves corresponding to the values of the coupling parameter $F=1$
($F=0.1$) in each case.  (b)
Discovery region (which lie above the curves)
for single leptoquark production
as a function of the leptoquark mass for the values of integrated
luminosity as indicated.}

\noindent
{Fig. 13.  $95\%$ C.L. indirect search limits on the leptoquark mass and
coupling $\kappa=2F$ from $\epem\to q\bar q$, assuming $sqrt s=500\gev$ and
${\cal L}=5\infb$.  The discovery region lies above the curves.}

\noindent
{Fig. 14. (a) $\sigma(e^-e^-\to W^-_LW^-_L)$ as a function of the heavy
neutrino mass for $\sqrt s=0.5\, (1)\tev$ corresponding to the solid
(dashed-dotted) curve.  The cross section should be rescaled by 4 powers of
the neutrino mixing angle.  (b)  Total cross section for $e^-e^-\to W^-_RW^-_R$
as a function of the CM-energy for various values of the mass of the heavy
right-handed neutrino $\nu_2$.  (c)  $W^-_LW^-_R$ production as a function of
the heavy neutrino mass $m_N$ with $\sqrt s=1\tev$.  This result must be
rescaled by two powers of the mixing angle ratio $\theta/0.01$.}

\noindent
{Fig. 15.  Total cross section for $e^-e^-\to e^-e^-$ in the presence of a
500 GeV dilepton as a function of the CM-energy.}

}
\end{document}